\documentclass[9pt,twocolumn,twoside,lineno]{pnas-new}

\templatetype{pnasresearcharticle} 

\newcommand{\new}[1]{{\color{orange}{#1}}}

\title{A systems framework for remedying dysfunction in U.S. democracy}

\author[a,1]{Samuel S.-H. Wang}
\author[b,1]{Jonathan Cervas} 
\author[c]{Bernard Grofman}
\author[d]{Keena Lipsitz}

\affil[a]{Neuroscience Institute, Princeton University, Princeton, New Jersey USA. ORCID: 0000-0002-0490-9786}
\affil[b]{Institute for Politics and Strategy, Carnegie Mellon University, Pittsburgh, Pennsylvania USA. ORCID: 0000-0001-9686-6308}
\affil[c]{Department of Political Science, University of California, Irvine, California USA. ORCID: 0000-0002-2801-3351}
\affil[d]{Queens College and the Graduate Center, City University of New York, Flushing, New York. ORCID: 0000-0002-9192-836X}
\affil[1]{Electoral Innovation Lab}

\leadauthor{Wang} 

\significancestatement{We suggest that a systems-based theory provides a natural vocabulary for evaluating the effectiveness of electoral reform in the United States. Many principles identified by political science research can be described in the language of engineering and physical sciences, including mechanisms that interact in a complex manner to drive representational outcomes. Such translation eases the way for modelers to investigate political dysfunction, as well as understand the effectiveness and robustness of proposed reforms.}

\authorcontributions{S.W., J.C., B.G., and K.L wrote the paper.}
\authordeclaration{S.W. is the founder of the Electoral Innovation Lab which conducts rigorous empirical research on democratic reforms. J.C. is a research associate at the lab. J.C. is consultant to the Pennsylvania Legislative Reapportionment Commission. Other authors declare no competing interest. B.G. is the Peltason (Bren Foundation) Chair at UCI. S.W. is supported by Arnold Ventures, Schmidt Futures, the Marilyn J. Simons Foundation, and Educational Ventures.}
\equalauthors{S.W., J.C., B.G., and K.L. contributed equally to this work.}
\correspondingauthor{To whom correspondence should be addressed. E-mail: sswang@princeton.edu}


\begin{abstract}
Democracy often fails to meet its ideals, and these failures may be made worse by electoral institutions. Unwanted outcomes include elite polarization, unresponsive representatives, and the ability of a faction of voters to gain power at the expense of the majority. Various reforms have been proposed to address these problems, but their effectiveness is difficult to predict against a backdrop of complex interactions. Here we outline a path for systems-level modeling to help understand and optimize repairs to U.S. democracy. Following the tradition of engineering and biology, models of systems include mechanisms with dynamical properties that include nonlinearities and amplification (voting rules), positive feedback mechanisms (single-party control, gerrymandering), negative feedback (checks and balances), integration over time (lifetime judicial appointments), and low dimensionality (polarization). To illustrate a systems-level approach we analyze three emergent phenomena: low dimensionality, elite polarization, and anti-majoritarianism in legislatures. In each case, long-standing rules now contribute to undesirable outcomes as a consequence of changes in the political environment. Theoretical understanding at a general level will also help evaluate whether a proposed reform's benefits will materialize and be lasting, especially as conditions change again. In this way, rigorous modeling may not only shape new lines of research, but aid in the design of effective and lasting reform.
\end{abstract}

\dates{This manuscript was compiled on \today}
\doi{\url{www.pnas.org/cgi/doi/10.1073/pnas.XXXXXXXXXX}}

\begin{document}

\maketitle
\thispagestyle{firststyle}
\ifthenelse{\boolean{shortarticle}}{\ifthenelse{\boolean{singlecolumn}}{\abscontentformatted}{\abscontent}}{}


\dropcap{I}n the face of record distrust and dissatisfaction with respect to American political institutions, interest in reform measures has exploded \cite{Lessig2019, Daley2020, Drutman2020}. These reforms are aimed at addressing a range of ills, including political polarization, governmental gridlock, and the failure of a majority of legislators to be elected by a majority of voters, among others. Ideas for reform range from ranked-choice voting to redistricting reform to changes in the judiciary.
 
No matter how well-intended any reforms may be, their consequences are not always easy to predict. Research on the outcomes of reform often examines specific cases without extracting general principles. We suggest that it will be useful to develop a framework which is both consistent with empirical research and extends that research to generate a theory of applied reform. A perspective based on complex systems synthesizes the efforts of reformers and academics, and helps identify steps with the most leverage to achieve particular goals, while avoiding undesirable, unexpected outcomes.
 
 \new{Political scientists have argued that the discipline needs to use methods that, "...combine sensitivity to causal complexity and causal effects with aspirations to draw out implications about social processes that transcend a single social setting..." \cite[p. 169]{Pierson2011}. Such arguments are a response to the dominance of regression and rational choice analysis in the discipline. Scholars using these methods tend to decontextualize processes and human interaction in search of parsimonious theories, identifying linear trends and fixed points. Such theories have been used mainly retrospectively to describe past events. Dissatisfaction with the ability of these methods to model complexity has led some researchers to look elsewhere. For example, Nobel laureate Elinor Ostrom drew on models from natural sciences to create a framework for studying socio-environmental systems that integrates the analysis of systems resources, users, and governance \cite{ostrom2007diagnostic, ostrom2009general}. Pierson has suggested that traditional methods have difficulty incorporating dynamic processes \cite{Pierson2011}. Game-theoretic models seek static equilibrium points, but these are difficult to find once too many factors are introduced. To account for real-life political events and political behavior, a more detailed analysis would incorporate individual behavior, various contexts, and specific mechanisms (institutions, rules, norms, etc.).}
 
\new{The advent of complex systems analysis, as well as widely available computing power, make it possible to formulate and interrogate detailed models. \new{Simulation methods have been driven past successes in political science research \cite{Axelrod1984, Axelrod1998}. Recently, Laver \cite{Laver2020} has used agent-based simulations to model polarization. In this issue of \textit{PNAS}, Leonard et al. \cite{Leonard_et_al_2021} investigate mechanisms driving elite polarization in the United States and demonstrate a key role for self-reinforcing elite processes. Such modeling may also be useful in formulating interventions: Axelrod et al. \cite{Axelrod2021} use detailed modeling to ask whether policy interventions can stop runaway feedback loops that lead to extreme polarization. It is even possible to construct mathematical models that go beyond political science to address political-societal instability more broadly using a variety of historical data \cite{Turchin2016}.} By drawing on a complex systems approach, the study of applied reform we propose can provide academics and reformers with a better understanding of how political systems work -- and how to rescue them from breakdown.}
 
\new{Creating such a research framework requires} bridging the gap between problems of democracy and the research toolkit of those in the natural sciences. The articles in this issue of \textit{PNAS} illustrate a variety of approaches that draw upon expertise in complex systems and other computationally and modeling-intensive disciplines such as mathematical biology. But despite the keen interest of these researchers, they face the task of making a meaningful contribution in a discipline with its own mature traditions, political science. The problems of democracy are sufficiently important to warrant an attempt at translation.

We hope our effort to render the U.S. political system in a mathematical complex-systems framework will encourage participation by scholars of the natural sciences. Specifically, we hope it will allow three outcomes: (a) the ability to build models that reproduce political phenomena from realistic parts, (b) the creation of simulations to explore alternative scenarios, and (c) the design and evaluation of interventions that may improve the function of democracy. These goals are analogous to those of engineers -- to understand a system of many parts well enough to make repairs or improvements. 

\new{The task of repairing American democracy could not be more urgent. Political scientists have found that democracies can decay, backslide, and even die \cite{Huntington2012, Hasen2020ElectionMeltdown, levitsky2018democracies, Mettler_Lieberman_2020}, reminiscent of the potential for natural systems to undergo sudden change such as avalanches and ecological collapse}. \new{Indeed, articles in this issue suggest that certain dysfunctions of American democracy, such as elite polarization, may have already passed a point} \new{of no return} \new{\cite{Leonard_et_al_2021}. Every avenue of investigation into the processes unfolding around us must be used, especially those that are capable of modeling complexity. It is an all-hands-on-deck situation.}
 
\section*{A Complex Systems Approach}
 
In engineering or biology, one encounters complex systems of interacting parts. Systems may be designed, such as a power grid or a mechanical clockwork, or naturally occurring, such as animal population dynamics or the evolution of new species. In both cases, a full understanding of a system’s behavior requires understanding individual rules, network interactions, and the effects of exogenous factors. 
 
A complex-systems approach to the study of politics seeks to account for emergent behaviors that arise from such combination of features. Broadly, a complex system may have history-dependence, undergo sudden transitions (criticality), show nonlinear relationships, retain a memory of past events (hysteresis), and be nested (components may themselves be complex systems). A complex-systems approach can help to identify interactions, understand long-term dynamics, and recognize ways in which changes in any part of the system can lead to unintended consequences when new conditions arise. Political scientists often write about these features but use different terminology.
 
One of the difficulties of understanding reforms is that the same upstream cause does not always have the same downstream consequence. For example, the stress hormone response may mobilize energy to support a temporary need such as normal exertion, yet also lead to dysfunction when chronically activated. Similarly, the rules of representative democracy may lead to stable governance or severe dysfunction, depending on the circumstances. Coming up with a treatment for what ails democracy requires some understanding of how remedies may lead to an abatement of symptoms—and whether the treatment is beneficial in the long run, or interferes with other treatments. In short, a system-level view can help us understand how conditions are filtered through rules and may help craft reforms that are effective in a lasting manner.
 
Below, we discuss some of the elements of a complex system and how they apply to a democracy, such as the United States. We focus on how rules and institutions can work to encourage (or discourage) three dysfunctional emergent properties: unidimensionality, polarization, and anti-majoritarianism.

\subsection*{Background Conditions}

Political behaviors arise from a combination of designed and naturally-arising features that include institutions, demographics, and geographic variation. A central design challenge to any reform arises from the fact that formal institutions are both mutually embedded and embedded in wider society that undergoes continual change \cite{Starr2019}. 
In the United States, rules and institutions from 1790, when voters comprised white male landowners and slaveowners in a nation of four million, were not designed to address today’s governance needs. Moreover, existing rules and institutions may amplify background conditions that drive polarization. 

The decline of civic life in America and the pluralism it once nurtured has hastened a collapse of dimensionality in the system. Americans once enjoyed a rich associational life \cite{Putnam2000}. Non-political associations, such as labor unions, churches, and bowling leagues, were often cross-cutting, bringing people from different backgrounds into contact with one another, building trust and teaching tolerance. To be sure, these associations often did not do enough to bridge certain cleavages, particularly those pertaining to race, but they did preserve some higher dimensionality in the system. Many of these associations have disappeared or have converted from membership groups that encouraged civic participation to managed groups focused on lobbying, such as the National Rifle Association \cite{Skocpol2013}. In short, the groups that once structured a multidimensional issue space in the United States have collapsed.

Dimensional collapse has been driven over the last century \new{as} economic and demographic changes have made the United States more urban \cite{krugman1991geography}, diverse \cite{Frey2018}, and unequal \cite{Bartels2017, Schlozman_et_al_2018}. Over the last century, rural towns have declined from 50\% of the population to 20\% today. The Hart-Celler Immigration and Naturalization Act of 1965 helped bring the immigrant population from 5.4\% to 13.7\% of all inhabitants. 
\new{In addition, the Voting Rights Act of 1965 dramatically increased Black enfranchisement in the Deep South, adding to the diversity of its electorate \cite{Alt1994, Grofman_Davidson_1994}.} Finally, wealth has become concentrated in the hands of the super-wealthy with only modest gains at the middle of the income distribution and losses of working-class jobs \cite{Piketty2003}. The connection between these profound changes and political dysfunction is not obvious until one considers what happens when the major cleavages they have produced -- rural/urban, white/non-white, rich/poor -- begin to align. Ideology has joined white, rural voters with the wealthiest Americans under the banner of the Republican Party while the Democratic Party has become a coalition of the rest \cite{GrossmanHopkins2016}. Taking the United States as a complex system, these developments can be understood as an intensification of political conflict along a single axis of variation. In order to fit into two political parties, these multiple issue conflicts have become projected into a single dimension \cite{Duverger1954, Downs1957}; see also \cite{Hinich1994, TaageperaGrofman1985}.
 
When the two parties are closely divided in strength, as they have been for the last 20 years, substantial advantage comes from small increases in support \cite{Lee2016, Fiorina2017a}. Political competitiveness can therefore amplify unidimensional conflict. Close divisions, especially when single-party control is seen as a possibility, exacerbate polarization \cite{Lee2016} leading to a more confrontational tone of ``constitutional hardball'' \cite{Tushnet2004} in which governing norms are broken and rules are bent, especially in the service of gaining advantage. Incivility and norm violations by one side are answered in kind when the other side returns to power. Violations such as gerrymandering, difficult judicial appointments, and abuses of election rules and laws foster a contentious and distrustful environment that further exacerbates animosity. \new{In a disturbing trend, polarization has recently developed an anti-establishment populist component that is correlated with acceptance of political violence \cite{Uscinski_et_al_2021}. Future political stability may depend on whether this component becomes the dominant axis of politics or can be kept orthogonal.}
 
\subsection*{Rules, institutions, and emergent political phenomena}
 
The rules of democracy, and the political institutions they create (e.g. Congress, the presidency, and state legislatures), function within the complex system described above. In contrast with biological evolution, whose rules change slowly through generational inheritance guided by survival fitness, rules pertaining to politics are created by interests with a variety of goals, which often work at cross purposes \cite{Starr2019}. In addition, political rules exhibit strong inertia because political actors will fight to preserve those they find beneficial and existing rules constrain future moves \cite[][p. 16]{Schickler2001}. Thus, rules can rarely be rewritten wholesale. Instead, new ones are layered on top of old ones creating ``disjointed’’ institutions and processes \cite{Schickler2001}. 
 
Democracy’s rules can interact with broader societal forces in a variety of ways. First, existing rules can act upon changing background conditions in ways that create new problems or reveal hidden problems in the design of government that were previously latent. Rules can exacerbate certain properties of the system (\textit{amplification}) or entrench existing states, making change difficult (\textit{hysteresis}). Legislators and judges can alter the rules to enhance or reduce such consequences, a positive or negative \textit{feedback} effect. Third, rules can serve as \textit{integrators over time}, converting temporary shifts in power to long-lasting changes that resist change or mitigation. Finally, the combined effects of these mechanisms can generate nonlinear dynamics, leading to rapid acceleration of a particular characteristic of the system beyond a \textit{critical threshold} \cite{Leonard_et_al_2021}, leading to sudden change that is not predicted by gradualist analysis, i.e. revolution or collapse.
 
\textit{Criticality} is an especially important emergent property: a collective property with no direct explanation in terms of the properties of individual components. Other examples of emergent properties include flocking and schooling, the brain, and cities. Viewed from a systems perspective, many dysfunctions of democracy are emergent properties, such as gridlock, antimajoritarianism, and notably in modern times, polarization of institutions and leaders. In political systems, processes that lead to criticality are often slow-moving, having a negligible impact until they reach a threshold \cite{Pierson2011, Turchin2016}. For example, gradual demographic change can account for the occurrence of revolutions \cite{Goldstone2016}.
 
Critical transitions in a variety of phenomena may have early warning indicators that include slowing down, increased variance, and skewness \cite{Hagstrom2021}, and may be seen in phenomena as diverse as market crashes and avalanches \cite{Bouchaud2013}. Of particular note as a forerunner to critical transitions is a collapse of dimension \cite{Hagstrom2021}, calling to mind the current low-dimensionality of U.S. politics. Mapping these concepts onto the political world in a rigorous manner may be of practical value in detecting and even averting future disruptions. For example, several papers in this volume argue that polarization trends in this country exhibit the property of criticality \cite{Leonard_et_al_2021, Szymanski2021, Vasconcelos2021}. The following properties of a complex system may be implicated in the nonlinear processes producing such phenomena.
 
\subsubsection*{Interaction of rules and background conditions} Existing rules and institutions can exacerbate or mitigate drivers of particular properties while remaining nominally unchanged. As background conditions change, rules can start to produce outcomes that are at odds with fundamental goals in a democracy. For example, if a country with a single-member plurality district system like the United States has voters who are relatively ethnically and economically homogenous, as was the case during the early republic, it matters less if districts have an unequal population size. As a population becomes more heterogeneous, however, and particular communities become more concentrated in certain areas, the equality of district populations becomes more important for representation. While state legislatures were directed to engage in decennial redistricting by the Constitution, they failed to do so until compelled by the Supreme Court in the 1960s. The lack of enforcement allowed powerful rural state legislators to maintain their power by ignoring population growth in their states’ cities through ``silent gerrymanders’’ in which the most populous district in a legislative chamber could have a population a hundred times larger than the least populous district. Thus, changing background conditions produced gross distortions in representation, introducing a stress in the system that could only be ameliorated by a change in rules: the requirement that districts have equal population size \cite{Ansolabehere_Snyder_2008}.
 
\subsubsection*{Feedback} In addition, rules and institutions can affect certain properties by creating either positive or negative feedback. For example, when legislators are in charge of redrawing their own district lines, the resulting protection of incumbents from voter opinion, termed gerrymandering, forms a positive feedback loop that reinforces the legislature’s power. Conversely, one branch of government may block another from acting, forming an inhibitory feedback step. Nominally, American systems of government have an unusually high number of veto points, including presidential or governor’s vetoes of legislation, limits imposed by the judiciary, and even rules that prevent a whole body (the Senate) from acting if it fails to have a supermajority \cite{Krehbiel1998a}. These inhibitory steps act to counteract branches of government acting alone. If enough inhibitory steps are removed, it becomes possible to enter a regime of runaway positive feedback. But too much inhibition can also paralyze the policy process. 
 
\subsubsection*{Integration over time} Integration occurs in a political system when a particular rule or institution entrenches political power, fixing a particular state. For example, federal judges are appointed by the President and serve for life. These judges then rule on cases that may affect the power of Congress and later Presidents. By outlasting the President’s tenure in office, the judiciary effectively integrates Presidential influence over time.
 
The temporally-integrative nature of federal judiciary power increases incentives for legislators and the President to intensify polarized conflict over judicial appointments. Indeed, any institutional rules that would give a long-lasting advantage provide an incentive to engage in hardball tactics \cite{Tushnet2004}. As judicial appointments take on outsized importance, they in turn become an issue that motivates polarization, thus creating a feedback loop. For example, confirmation of judicial appointments used to require the approval of both of an appointee’s home-state Senators as well as supermajority support on the floor of the Senate; now, confirmation rules have been changed to require only bare-majority support and votes routinely follow partisan lines. In this case, the new rule both arises from polarization and acts in positive feedback to drive further polarization.
 
\section*{A Systems-Driven Framework For Reform}

Many problems of democracy seem intractable. Some proposed reforms require altering the Constitution, a step which requires ratification by a supermajority of states and is unsuited to current, closely divided times. At least one — abolishing the Senate — would require a constitutional convention. Other changes may be implemented, however, by passing a federal law, state law, or citizen initiative. In this section we focus on the latter. 
If existing rules lead to unwanted new consequences due to changing conditions, then well-designed new rules may be able to reduce or reverse those consequences. We describe several emergent dysfunctions in U.S. democracy, identify likely causes, and review potential reforms that can realistically be made law in the near future. We will then consider the known effectiveness of the reforms, and outline research directions to test their robustness and durability to varying conditions. Our proposed approach to research can both reveal anticipated short-term improved function, as well as long-term pitfalls. In each case, changes are consistent with the current system of rules, constituting only additions or modifications that are arguably consistent with the Constitution.
 
\subsection*{Emergence 1: Loss of dimensionality}
The loss of dimensionality in the system has affected the attitudes of voters and elected officials, contributing to political polarization, which is associated with three central features: (a) bimodality of opinions on issues and a high standard deviation; (b) increased cross-issue correlation, so that there are fewer voters or legislators who are cross-pressured in their choice among policy platforms \cite{Gould2022}; and (c) the sorting of voters and legislators into parties so that the parties are increasingly homogeneous with respect to any given issue and in overall ideological terms.

At the level of \new{public officials \cite{mccarty2016polarized} and activists \cite{Evans2003}} there is strong evidence for all three features of polarization. At the mass level, there is compelling evidence for sorting, strong evidence for growing cross-issue correlations, and equivocal evidence for strong polarization on individual issues and beliefs \cite{Fiorina2017a}. Even if voters are not ideologically polarized, there is abundant evidence that they are polarized affectively, i.e., they dislike elected officials and even voters from the opposing party simply because of their partisan identification \cite{Iyengar2019}. The electorate is also polarized by racial attitudes \cite{tesler2010obama}. When voters are unwilling to punish extreme candidates or incumbents by not voting for them, a crucial mechanism for moderating polarization in the system is \new{lost \cite{Grimmer2013}.} 

\new{Low dimensionality may be driven by change in the nature of news media. Nationally, the once-dominant ABC-CBS-NBC triad has been replaced by increasingly divergent news sources that silo political views into opposing camps. At the same time, the disintegration of local journalism has radically reduced the availability of information that may differentiate regions of the nation politically \cite{Darr2018, Peterson2019, Hopkins2021}. These changes enable extremism of political debate at a national level \cite{Merrill2014, Zelizer2020}.}

Can dimensionality be increased through interventions? Encouraging voters to make choices based on some dimension that cuts across the urban/rural and white/nonwhite cleavages previously mentioned could be a partial solution. This is not easy given how nationalized our politics have become \cite{Jacobson2013}. Belying the old saying, “all politics is local” \cite{ONeill1994}, modern candidates campaign almost exclusively on the national party platforms. 

Reforms that force elected officials to address state and local issues, as well as deliver on those campaign promises, 
\new{could} increase dimensionality. \new{One such reform might include} strengthening state parties. 
The pluralistic nature of state parties used to be a mitigating force against polarization \cite{Pierson2020a}. Strengthening state parties can build local independence and make the parties more heterogeneous at a national level \cite{McCarty2015}. State-party independence depends in part on campaign finance. In state legislatures, party money tends to flow to more moderate candidates while individual and interest group money tends to flow to more extreme members \cite{LaRaja2015}. The McCain-Feingold campaign finance reform weakened state parties, as has the rise of independent expenditures. La Raja and Schaffner suggest that the balance may be shifted back toward local interests by removing most limits on state party fundraising in addition to subsidizing them in other ways, for example with a publicly-financed matching program that matches in-state donations at a higher rate than out-of-state donations. The desired effect would be to strengthen state parties so they can withstand pressure from the national parties. 

However, it is not known how effective such a reform would be. Understanding the consequences of local campaign finance reform could be attained through local experimentation, which would take a period of years. An alternative would be the use of methods to simulate motivations and actions. Understanding the source of orthogonal forces, how they are currently incorporated into the American system, and how they are translated into power and policy by voting rules and other institutions, are important areas for future work. 
 
\subsection*{Emergence 2: Polarized elites}

Polarized elected officials are often pulled to extremes by interest groups and activists who threaten to ``primary’’ them by funding more extreme candidates if the elected officials do not pass their desired policies. Party activists also ensure that the national parties include more extreme policy proposals across an ever-expanding set of issue areas, promoting “conflict extension” \cite{Layman2002} in their platforms. At the same time, it is generally recognized that contests for the U.S. Congress have become increasingly nationalized, so that the destiny of a candidate for federal office is tied to national political forces \cite{Merrill2014, Abramowitz2016, Carson2020, Jacobson2019}. If candidates of a given party cannot stray far from the national party positions that are at the extremes, then a district’s median voter cannot be courted in the idealized fashion described by formal theory \cite{Downs1957}. As interest groups and activists achieve more success, the emergence of ``teamsmanship’’ encourages unaligned interest groups to choose a side and intensifies conflict \cite{Lee2009a}. Studies suggest these sources of feedback have pushed or are close to pushing elite polarization over a critical threshold \cite{Leonard_et_al_2021}.

Can election reforms reverse or at least mitigate the processes that drive elite polarization? Existing rules provide a target for change. \new{In a low-dimensional system, one potentially stabilizing force is the election of candidates who are close to the median voter. Modern democratic theory \cite{Downs1957} posits that the median voter ought to be the one whose policy preferences are realized. Yet under current rules, the decisive point on the spectrum (i.e. the pivotal voter) is often far from the median.} The most common rule for electing legislators in the United States, however, is a two-stage election, with a primary election to determine nominees, followed by a general election. Most often, both elections are determined by a plurality vote, also known as first-past-the-post. Under this rule, winners are only guaranteed to require majority support when there are two candidates. Legislators with extreme ideological positions can win office either by winning an extreme bloc of votes in the primary, and/or by splitting the vote among multiple opponents in the general \cite{thomsen2017opting}. This system is known to generate representational distortion by rewarding off-median winners in party selection mechanisms, and by failing to elect Condorcet winners, i.e., candidates who would win every one-on-one pairing with individual opponents \cite{Grofman_Feld_2004}. 
 
\subsubsection*{Ranked-choice voting} Instant-runoff voting (often called ranked-choice voting, RCV) is designed to drive outcomes toward winners who are acceptable to the majority. By successive elimination of poor-placing candidates, the field is reduced to two finalists, one of whom is elected by a majority of all voters expressing a preference between them. This method can be used in either primaries or the general election.

Ranked-choice voting has recently been adopted in a variety of jurisdictions, including Congressional and state elections in Maine, mayoral primaries in New York City, and statewide elections in Alaska. Alternative voting rules such as RCV can elect Condorcet winners more often than a plurality rule. They may also allow voters flexibility to show support for a long-shot candidate without hurting the chances of a less-preferred candidate with a higher likelihood of electoral success. Such rules may also temper extremism by providing an incentive for candidates to appeal to the median voter and reduce negative campaigning. \new{On average, RCV does not appear to favor either major political party \cite{Cervas2021RCV}.} However, new voting rules bring burdens, including a remaining risk that a centrist candidate can sometimes be eliminated in early rounds \cite{Fraenkel2004,Fraenkel2006}, and an increased workload on voters to rank many candidates. Jurisdictions that use RCV now span a range from towns to entire states, providing a rich source of natural experiments for probing differing levels of voter polarization, socioeconomic stratification, and engagement. 

Research into alternative voting rules appears ready to transcend the examination of individual examples. \new{In theory, ranked-choice methods can reduce polarization and improve representation\cite{Cox1997}. But existing studies, which explore only a handful of particular natural settings, have been equivocal \cite{Neely_McDaniel_2015, Burnett_Kogan_2015, Donovan_et_el_2016, John2018, Santucci2018, Santucci2021}. The fact that these studies span a variety of conditions (city elections, federal elections, strong/weak parties) and voting rules (instant-runoff, top-two, top-four, with and without primaries) raises the possibility that conclusions to date are based on inadequately stratified data. Simulation-based approaches address this problem by allowing the rapid exploration of a larger parameter space. In this way, examination of specific cases limits the pace at which lessons may be obtained to drive improvements in practice.
 
Synthetic methods can address the stratification problem by allowing a wide range of conditions to be explored artificially. A first step might be to identify the dimensionality and heterogeneity of voter populations in existing ranked-choice elections using cast vote records, allowing the production of a generalized model of voter preference and behavior. Such a model could be combined with voter and candidate strategies to understand the effects of different voting rules and local conditions. Such a generative model would draw upon existing voting data, opinion surveys, cognitive science, and game theory. This research program could also identify ways to reduce undesired effects such as ballot exhaustion, decreased turnout, and shutting-out of a political party. The emerging general understanding might achieve a rate of discovery rapid enough to guide the design of new and better reforms.}

\subsubsection*{Eliminating closed primaries} A first-past-the-post primary election can feature many candidates, in which case the nominee can win with considerably less than half of the vote. Because of this, a cohesive minority, including one composed of extremists, may potentially determine one or both nominees \cite{Bishin2009}. In principle the power of a small number of voters is enhanced further by the fact that turnout is limited to party members who are engaged enough to vote. Through primaries, rank-and-file party members can reward loyalty to the party’s issue positions and tone, thus perpetuating polarization.
 
Primary election rules would seem then to be a ripe ground for reform. For example, the selection rule could be changed to a ranked-choice election, or voters who are not party loyalists could be permitted to participate. Possible reforms include the opening of primaries to allow nonmembers to vote, merging party primaries to a single nonpartisan top-N system (followed by an N-candidate general election), or the introduction of new selection rules such as RCV or approval voting. 
 
Although these reforms are logically appealing, their success in electing more moderate candidates is mixed so far. For example, open participation in primaries has been reported to cause minor change or no change in legislator voting patterns \cite{Mcghee_et_al_2014, McGhee2017b, Ahler2018}. These negative findings may face the difficulty of stratification: the effectiveness of a reform may depend on particular local factors, yet examining particular situations may lack statistical power. As more examples become available, statistical power can improve. Indeed, using similar methods as a negative report focusing on state legislatures through 2010 \cite{Mcghee_et_al_2014}, a more recent study encompassing Congressional seats through 2020 \cite{Grose2020_JPIPE} found a moderating influence from California’s top-two primary system. A simulation-based approach may be useful in identifying the factors that determine the effectiveness of various primary election reforms.

\subsubsection*{Emergence 3: Antimajoritarianism}
A recent feature of national politics is antimajoritarianism, the tendency for a majority of voters to fail to take control of the House, Senate, or Presidency. This tendency has been highlighted by the closeness of national elections since the 1990s, and arises from a combination of demography, geographic variation, and redistricting. 

As mentioned, polarization has a spatial dimension with an urban-rural divide emerging in the last half century \cite{Chu2021}. By 1968, the distribution of partisan preference showed a strong skew, with high-density states being more Democratic, and Republican tendencies in low-density states. Because Democratic-leaning states have higher population density and higher populations, representation in the Senate has taken an antimajoritarian turn. The urban-rural-derived partisan advantage arises from the fact that the Senate assigns two Senators per state irrespective of population \cite{Dahl2003, Cervas2020}. Although the Senate is split evenly after the 2020 election, the fifty Democratic Senators represent 41 million more people than the fifty Republican Senators. 

From 2002 to 2020, the fraction of the national fifty-state vote needed to attain a 50-50 split has averaged 48.0\% Republican; in other words, structural minoritaritan rule 
\new{with a 2.0-percentage point bias of popular vote relative to representation \cite{Wang_GH_2021}. A similar bias for selecting the President has grown in the last two elections \cite{Cervas2020, Erikson2020_PNAS}. One suggested Senate reform, the granting of statehood to Washington D.C., would reduce the gap to 1.7 percentage points. Eliminating the bias entirely would require admission of at least six more states, comparable to the wave of admissions orchestrated in 1889 and 1890 to give advantage to late-19th-century Republicans \cite{Richardson2020}.} 

The same density-partisanship correlation persists at a county and precinct level, affecting congressional and state legislative districts. Members of Congress and most state legislators are elected from nonoverlapping districts whose boundaries must be redrawn every 10 years, following the decennial census. These districts must be of approximately equal population within each state and cannot cross state boundaries. A partisan advantage can accrue when one party controls redistricting and “packs” voters of the opposing party forcing them to waste votes. Because Democrats tend to live in more densely populated areas, they are easier to pack, giving Republicans a natural advantage. The representational distortions that emerge from such partisan gerrymandering can be quite large \cite{Wang2016_SLR}.

\begin{figure*}
\centering
\includegraphics[width=.8\linewidth]{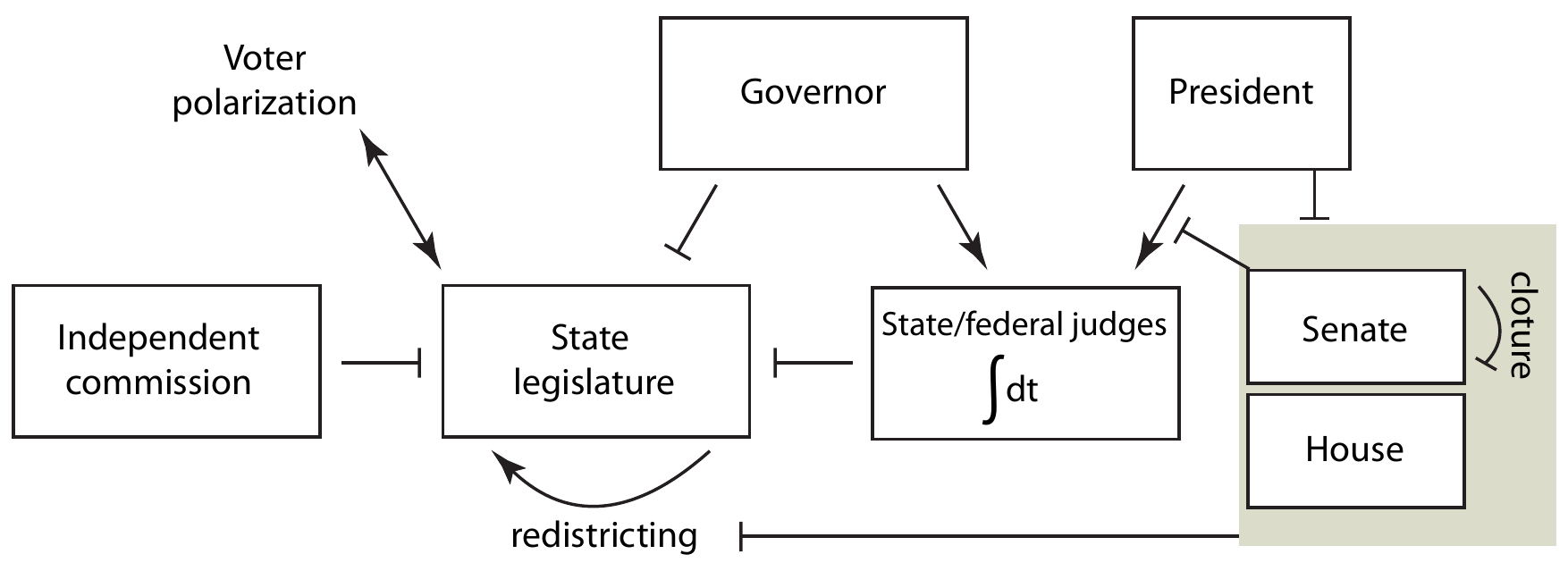}
\caption{A systems-level control diagram for redistricting. Feedback mechanisms described in this article are diagrammed in terms of directional control. Arrows indicate positive feedback, blunt symbols indicate negative feedback, and persistent effects are indicated with an integral sign.}
\label{fig:diagram}
\end{figure*}

Can representational distortions in legislatures and the U.S. House be reduced through reform? The diagram in Figure \ref{fig:diagram}, which illustrates the legislative process driving redistricting, illustrates how the redistricting process can be understood from a complex systems perspective. In most states, redistricting is like other legislation: it is done by the legislature by majority vote and must be approved by the governor \cite{Levitt2020}. Such a diagram may be useful for making the interactions of legislation accessible to those in the sciences. An engineering or biology-style interaction diagram reveals positive or negative feedback steps and integration over time which together can account for partisan gerrymandering. 

\textbf{Positive feedback}, indicated by an arrow that circles back toward its origin, has been facilitated in recent decades by increasingly reliable voter behavior and convenient redistricting software. It is possible to generate districts that perform in a predictable manner for an entire decade. \textbf{Inhibitory steps}, indicated with arrows ending with a blunt end, indicate ways in which one branch of government may block another from acting. If enough inhibitory steps are removed, the resulting runaway positive feedback allows an unchecked legislature to gerrymander itself into power and stay there for ten years, all the way to the next cycle. 
 
The Supreme Court has weakened protections for partisan and racial fairness in recent years (\textit{Shelby County v. Holder}, 570 U.S. 529; \textit{Rucho v. Common Cause}, 588 U.S. \underline{\quad}) and further erosion may arise in a Supreme Court case (\textit{Brnovich v. Democratic National Committee}) pending at the time of this writing (June 2021). In short, both racial and partisan gerrymanders are now easier to commit than in 2010, effectively removing one of the inhibitory feedback steps in Figure \ref{fig:diagram}.
 
Reform can be accomplished by shifting the power of redistricting away from the legislature and to a non-partisan or bipartisan commission, or potentially by establishing explicit neutral (good government) or fairness criteria that can be enforced by courts. The potential for change in rules varies by state, especially in terms of whether the state allows for use of the citizen initiative. During the past decade, initiatives have taken redistricting away from the legislature and put it into the hands of a commission and the number of states with commissions has grown, though many of these commissions are only advisory. Independent redistricting commissions are currently considered to be constitutional; the recent rightward turn of the court indicates that the use of such commissions for federal redistricting may now be in question.

\section*{Future research}

Our aim in this paper has been to encourage those interested in complexity to study how political systems produce emergent properties like polarization. Modeling dynamical processes in political systems is daunting due to the complexity of cognitive and social processes, but the articles in this issue illustrate that it can be done and in a manner that produces important insights. Political scientists are already familiar with many of the dynamical processes discussed above and have rich literatures on many emergent properties of interest. For example, those who study how political institutions develop understand that they often show nonlinear behavior, can reach critical thresholds for change, and may contain amplifying feedback loops. 

However, questions in political science can take years to resolve, for reasons ranging from waiting for multiple elections and natural experiments to take place, to the pace of peer review and academic dialogue. This task is made more difficult by the rapid timescale and the possibility of criticality in U.S. politics. Quantitative modeling may accelerate discovery by allowing specific observations to be generalized to a useful framework for understanding when dysfunctions arise and when reforms may or may not work. \new{Dynamic simulations can extend the scope of quantitative analysis by predicting the efficacy of reforms -- or the consequences of inaction. Complex-systems modeling is useful in the same way regression is: it helps extrapolate to unknown situations. Regression methods are linear and address variations that occur within the range of known phenomena. Complex-systems-based methods open the possibility of addressing nonlinearities and instabilities under conditions that have not yet occurred. This addresses a ``fundamental problem of causal inference'' \cite{Rubin1974, Holland1986}} \new{in observational studies; the inability to} \new{test the efficacy of programs under novel conditions. Such an approach makes research more agile and can help bridge the gap between the rigor of scientific investigation and the time pressure faced by reformers.}

Complex-systems modeling may reveal how political collapse occurs — and how it may be averted. Close partisan divisions and political polarization have occurred together in the U.S. twice, first during the Gilded Age (1876-1896) and again in current times starting in 1994. In both eras, Presidential races have been decided with popular-vote margins of less than five percentage points in 5 out of 6 elections, and control of the U.S. House of Representatives has switched parties at least four times. In the first Gilded Age, belief in impending collapse was common. But what came next was the Progressive Era, a period of institutional reform and depolarization \cite{putnam2020upswing}. Can institutional reforms help guide us out of the new Gilded Age? An acceleration of discovery in the study of politics may help move U.S. democracy close to its stated ideals.

As fruitful collaboration arise between the natural and social sciences, we believe the natural scientists can be particularly helpful in three domains: (a) the ability to build models that reproduce political phenomena from realistic parts; (b) the creation of simulations to explore alternative scenarios without waiting for them to occur naturally; and (c) the design and evaluation of interventions that may improve the function of democracy. In this way, a complex-systems approach may be useful to help formulate a durable reform agenda on a practical time scale.

\acknow{The authors thank the staff of the Electoral Innovation Lab for assistance, Ari Goldbloom-Helzner for calculations, and participants of the Politics and Strategy Research Workshop at Carnegie Mellon University, the Political Polarization PNAS Conference, Jesse Tyler Clark, Lee Drutman, Christian Grose, Frances Lee, Simon Levin, and Rebecca Moss for discussion.}

\showacknow{} 

\bibliography{pnas-sample}

\end{document}